\documentclass{ifacconf}
\usepackage{amsmath} 
\usepackage{amsfonts}
\usepackage{amssymb}

\usepackage{algorithm}
\usepackage{algpseudocode}   

\newtheorem{definition}{Definition}
\newtheorem{remark}{Remark}
\newtheorem{lemma}{Lemma}
\newtheorem{theorem}{Theorem}
\newtheorem{proposition}{Proposition}

\newcommand{\qedbox}{\rule{1.2ex}{1.2ex}}
\newenvironment{proof}{%
    \par\noindent\textit{Proof:}\ }%
    {\hfill\qedbox\par}

\usepackage{graphicx}      
\usepackage{natbib}        
\begin{document}
\begin{frontmatter}

\title{Optimal Platoon Formation and Stable Benefit Allocation in Mixed-Energy Truck Fleets under Size Limitations\thanksref{footnoteinfo}} 

\thanks[footnoteinfo]{This research was supported in part by NSF under Grants CNS-2401007, CMMI-234881, IIS-2415478, and in part by MathWorks.}

\author[First]{Ting Bai},
\author[Second]{Xinfeng Ru},  
\author[First]{Andreas A. Malikopoulos}

\address[First]{Information and Decision Science Lab, School of Civil $\&$ Environmental Engineering, Cornell University, USA (e-mails: tingbai@cornell.edu, amaliko@cornell.edu).}
\address[Second]{School of Control Science and Engineering, Dalian University of Technology, Dalian, China (e-mail: rxf@mail.dlut.edu.cn)}

\begin{abstract}                
In this paper, we investigate cooperative platoon formation and benefit allocation in mixed-energy truck fleets composed of both electric and fuel-powered trucks. The central challenge arises from the platoon-size constraint, which limits the number of trucks permitted in each platoon and introduces combinatorial coupling into the search for optimal platoon formation structures. We formulate this problem as a coalitional game with bounded coalition sizes and derive a closed-form characterization of the optimal coalition structure that maximizes the fleet-wide platooning benefit. Building on this structure, we develop a type-based least-core payoff allocation scheme that guarantees stability within the coalition-structure core (CS-core). For cases in which the CS-core is empty, we compute the least-core radius to determine the minimal relaxation required to achieve approximate stability. Through numerical studies, we demonstrate that the proposed framework consistently achieves the highest total platooning benefit among all feasible formation configurations while providing stable benefit allocations that outperform existing baseline methods.
\end{abstract}

\end{frontmatter}

\section{Introduction}\label{Section 1}
Road freight electrification has emerged as a pivotal pathway for decarbonizing the transport sector in response to growing concerns over fossil fuel depletion and accelerating climate change~\citep{7904618,link2024rapidly,10932686}. Compared with conventional fuel-powered trucks (FPTs), electric trucks (ETs) offer substantial advantages, including higher energy efficiency, zero tailpipe emissions, and reduced maintenance and operating costs, contributing to a cleaner and more sustainable freight transportation system. To accelerate ET adoption, governments and private investors have introduced a range of supportive initiatives, such as purchase subsidies, carbon taxation policies, large-scale charging infrastructure development, and the development of optimal charging strategies~\citep{10147895}. As freight carriers progressively incorporate ETs into their fleets, traditional FPT fleets are gradually transforming into \emph{mixed-energy truck fleets} that contain both FPTs and ETs. This transition phase is expected to persist until road freight transportation is fully electrified. 

Truck platooning is an advanced cooperative driving technique in intelligent transportation systems and has been well developed over the past few decades~\citep{9976309,tsugawa2016review}. By leveraging vehicle-to-vehicle communication and automated cruise control, platooning enables trucks to travel in closely coordinated formations safely and seamlessly. The shorter inter-vehicle gaps and reduced aerodynamic drag, particularly for trailing trucks, yield multiple benefits, including increased road capacity, mitigated traffic congestion, enhanced driving safety~\citep{axelsson2016safety}, and reduced operational costs~\citep{10209062}. While the leading truck typically receives negligible energy savings, each follower can achieve a $10\%$-$20\%$ reduction in energy consumption~\citep{van2017fuel}, leading to substantial cost savings. In mixed-energy truck fleets, although platooning remains advantageous, the magnitude and nature of the benefits, such as energy consumption patterns and cost savings, vary significantly across vehicle types, making the formation of profitable platoons and stable benefit allocation scheme design more difficult.    

Forming platoons requires coordination among trucks, which can be achieved through various strategies, such as planning delivery routes to maximize the overlapping road segments~\citep{meisen2008data}, regulating trucks' velocity to merge into platoons~\citep{van2017efficient}, and scheduling waiting times at hubs along pre-planned routes~\citep{Wang2025programming}. These coordination mechanisms fundamentally rely on which trucks choose to cooperate and how the resulting benefits are shared among participants. In particular, identifying appropriate platoon partners determines the total platooning benefit that can be achieved, whereas the benefit allocation scheme governs how this benefit is distributed among individual trucks, shaping their incentives to cooperate. 

Although platoon formation and incentive-design schemes have been extensively explored for homogeneous FPT platoons, effective solutions for mixed-energy truck fleets remain limited. For instance, \citet{farokhi2013game} and \citet{johansson2019game} developed benefit-distribution models for the platoon matching problem, treating vehicles as strategic agents that independently adjust their departure times to maximize individual platooning gains. These interactions are formulated as a non-cooperative game, with Nash equilibrium used to characterize the stable platoon configuration. Similarly, \citet{sun2019behaviorally} addressed the problem by first maximizing the total platooning profit and then allocating benefits so that no vehicle has an incentive to deviate from its assigned platoon. However, these studies assume homogeneous vehicle characteristics, overlooking the imbalanced cost savings in mixed-energy fleets. Recently, ~\citet{chen2023cost} studied cost allocation among multiple carriers by optimizing vehicle departure times to minimize total operational cost while ensuring approximate core stability. Nevertheless, it considers only grand-coalition platooning while not accounting for the practical platoon size constraints.

This paper investigates the problem of optimal platoon formation and benefit allocation in a mixed-energy truck fleet comprising both FPTs and ETs. Due to safety and technological constraints, the size of each platoon is bounded, i.e., the number of trucks allowed to form a platoon is constrained. We address this problem within a game-theoretic framework by deriving an optimal platoon formation that maximizes the fleet-wide platooning benefit. Based on this formation, a payoff allocation scheme is developed to ensure the stability of the coalition structure. The major contributions of this work are:
\vspace{-2pt}
\begin{itemize}
\item We model the strategic interactions among trucks in forming mixed-energy platoons as a coalitional game and establish necessary and sufficient conditions for the existence of a feasible coalition structure under any fleet compositions and platoon size limits. 
\vspace{2.5pt}

\item We derive optimality conditions for identifying the platoon formation and leader selections that maximize the total platooning benefit attainable by the fleet. An efficient algorithm to construct the optimal coalition structure is developed accordingly.
\vspace{2.5pt}

\item Under the optimal coalition structure, we design a type-based least-core payoff allocation scheme that lies in the coalition-structure core (CS-core) whenever the CS-core is nonempty. In addition, it minimizes the least-core radius when the CS-core is empty, thus ensuring maximal formation stability.  
\end{itemize}
Finally, extensive simulation studies validate the theoretical findings and demonstrate the superior stability performance of the proposed framework compared to existing baseline methods. We note that this work extends our recent study in~\citep{Bai2025stable}, which addressed stable payoff allocation in mixed-energy truck platoons but did not take platoon size limitations into account. 

\emph{Notation:} Let $\mathbb{R}$, $\mathbb{R}_+$, and $\mathbb{N}$ denote the sets of real numbers, positive real numbers, and non-negative integers, respectively. The cardinality of a set $\mathcal{S}$ is denoted by $|\mathcal{S}|$. The ceiling operator $\lceil y \rceil$ gives the smallest integer greater than or equal to $y$, while the floor operator $\lfloor y \rfloor$ returns the largest integer less than or equal to $y$.

\section{Preliminaries and Problem Formulation}\label{Section 2}
This section formally presents the mixed-energy truck platoon formation problem within a coalitional game framework. First of all, some key concepts are introduced. 
\vspace{2pt}

\begin{definition}[Coalitional game] \label{Def1}
A coalitional game with transferable utility is a pair $(\mathcal{N},v)$, where \\
(1) $\mathcal{N}\!=\!\{1,2,\dots,N\}$ with $N\!\in\!\mathbb{N}$ is a finite set of players;\\
(2) $v\!:\!2^{\mathcal{N}}\!\rightarrow\!{\mathbb{R}}$ is a characteristic function, which assigns a real value $v(\mathcal{S})$ to each coalition $\mathcal{S}\!\subseteq\!{\mathcal{N}}$ with $v(\emptyset)\!=\!0$.\\
The game has transferable utility if the value $v(\mathcal{S})$ can be freely distributed among members of the coalition $\mathcal{S}$.
\end{definition}
\vspace{2pt}

\begin{definition}[Coalition structure] \label{Def2}
For a coalitional game $G\!=\!(\mathcal{N},v)$, $\mathcal{P}\!=\!\{\mathcal{S}_1,\dots,\mathcal{S}_P\}$ with $\mathcal{S}_m\!\neq\!{\emptyset}$ for $m\!=\!1,\dots,P$, is called a coalition structure (i.e., a partition of $\mathcal{N}$) if\\
\text{(a)} $\mathcal{S}_m\!\cap\!{\mathcal{S}_{\ell}}\!=\!\emptyset$, $\forall{\mathcal{S}_m,\mathcal{S}_{\ell}}\!\in\!\mathcal{P}$, $m\neq{\ell}$ (disjoint coalitions);\\
\text{(b)} $\bigcup_{m=1}^P\!\mathcal{S}_m\!=\!\mathcal{N}$ (complete coverage).
\end{definition}
\vspace{2pt}

\begin{definition}[Payoff vector] \label{Def3}
Given a coalitional game $G\!=\!(\mathcal{N},v)$ and a coalition structure $\mathcal{P}$, a vector $x\!=\![x_1,\dots,x_N]^{\top}\!\!\in\!{\mathbb{R}^N}$ is referred to as a payoff vector if\\
\text{(a)} $x_i\!\geq\!{v(\{i\})}$, $\forall{i}\!\in\!{\mathcal{N}}$ (individual rationality);\\
\text{(b)} $\sum_{i\in\mathcal{S}_m}\!\!x_i\!=\!v(\mathcal{S}_m)$, $\forall{\mathcal{S}_m}\!\in\!\mathcal{P}$ (coalitional efficiency).
\end{definition}

In the above definitions, each subset $\mathcal{S}\!\subseteq\!{\mathcal{N}}$ is known as a \emph{coalition}, and the set $\mathcal{N}$ is known as the \emph{grand coalition}. The characteristic function $v(\mathcal{S})$ represents the total value (i.e., payoff) that can be achieved by all players in $\mathcal{S}$ through cooperation in game $G\!=\!(\mathcal{N},v)$.
\vspace{2pt}

\begin{definition}[CS-core] \label{Def4}
Consider a coalitional game $G\!=\!(\mathcal{N},v)$ and a coalition structure $\mathcal{P}\!=\!\{\mathcal{S}_1,\dots,\mathcal{S}_P\}$. The CS-core is the set of all feasible payoff vectors $x\!=\![x_1,\dots,x_N]^{\top}\!\!\in\!\mathbb{R}^N$ such that no coalition external to $\mathcal{P}$ can profitably deviate. Formally,
\begin{align}
\text{CS-core}(\mathcal{P})&=\Big\{x\!\in\!\mathbb{R}^N \big|\sum_{i\in\mathcal{S}_m}\!\!x_i\!=\!v(\mathcal{S}_m),\ \forall{\mathcal{S}_m}\!\in\!\mathcal{P}, \nonumber\\
& \ \ \ \ \ \ \ \ \ \ \ \ \ \ \ \ \ \text{and} \sum_{i\in\mathcal{C}}x_i\!\geq\!{v(\mathcal{C})},\ \forall{\mathcal{C}}\!\subseteq\!{\mathcal{N}}\Big\}.\label{Equ.1}
\end{align}
\end{definition}

The concept CS-core extends the classical core of a coalitional game to settings where the grand coalition $\mathcal{N}$ is partitioned into multiple disjoint coalitions. Here, the condition $\sum_{i\in\mathcal{S}_m}\!\!x_i\!=\!v(\mathcal{S}_m)$, $\forall\mathcal{S}_m\!\in\!\mathcal{P}$, ensures that there is no wasted or unassigned value. The other condition $\sum_{i\in\mathcal{C}}\!x_i\!\geq\!{v(\mathcal{C})}$, $\forall{\mathcal{C}}\!\subseteq\!{\mathcal{N}}$ guarantees that no subset of players can deviate by forming a new coalition that gives every member a strictly higher payoff.
\vspace{2pt}

We are now ready to present the problem formulation. Throughout this paper, we consider a hub-based platoon formation game denoted by $G_p\!=\!(\mathcal{N},v)$, where trucks departing from an origin hub are grouped into platoons heading to the same destination hub. The set of players (i.e., trucks) is denoted as $\mathcal{N}\!=\!\{1,\dots,N\}\!=\!\mathcal{N}_e\!\cup\!{\mathcal{N}_f}$, where $\mathcal{N}_e$ and $\mathcal{N}_f$ represent the sets of ETs and FPTs, respectively. Each truck $i\!\in\!\mathcal{N}$ is associated with a type $T_i\!\in\!\{e,f\}$, indicating if it is an ET or FPT. The platoon size limitation is captured by $M\!\in\!\mathbb{N}$, with $2\!\leq\!{M}\!<\!{N}$.

Extensive field tests and experimental studies have indicated that follower trucks of the same type and load receive similar platooning benefits, while the benefit to the leader is significantly smaller and can be neglected~\citep{van2017fuel}. For this reason, we denote by $\epsilon_e,\epsilon_f\!\in\!\mathbb{R}_+$ the operational cost savings per travel time unit in platoons for an ET and FPT follower, respectively, and assume the leader obtains zero benefit. Since ETs generally incur lower cost savings than FPTs, we consider $\epsilon_e\leq{\epsilon_f}$.

The characteristic function $v(\mathcal{S})$ of game $G_p$ is defined as the maximum platooning benefit attainable by all trucks in the coalition $\mathcal{S}\!\subseteq\!{\mathcal{N}}$ via optimally assigning the leader role in $\mathcal{S}$. Specifically, let $S\!=\!|\mathcal{S}|\!=\!S_e\!+\!S_f$, where $S_e\!=\!|\mathcal{S}\!\cap\!{\mathcal{N}_e}|$ and $S_f\!=\!|\mathcal{S}\!\cap\!{\mathcal{N}_f}|$ denote the numbers of ETs and FPTs in $\mathcal{S}$, respectively. Then, $v(\mathcal{S})$ is denoted as
\begin{align}
v(\mathcal{S})&:=\max_{L\in\mathcal{S}}v(\mathcal{S},L)\nonumber\\
&=\begin{cases}
0,&\text{if } S\!=\!0,\\
\epsilon_e(S_e\!-\!1)+\epsilon_fS_f, &\text{if }1\!\leq\!{S}\!\leq\!{M}, S_e\!\geq\!{1},\\
\epsilon_f(S_f\!-\!1), &\text{if } 1\!\leq\!{S}\!\leq\!{M}, S_e\!=\!0,
\end{cases}\label{Equ.2}
\end{align}
where $L\!\in\!\mathcal{S}$ denotes the leader truck of the coalition (i.e., platoon) $\mathcal{S}$. Note that, given $\epsilon_e\!\leq\!{\epsilon_f}$, $v(\mathcal{S})$ is maximized by selecting an ET as the platoon leader whenever $S_e\!\geq\!{1}$.

In mixed-energy truck fleets, the fleet manager aims to maximize the total platooning benefit by presenting an optimal coalition structure $\mathcal{P}^*$ subject to the platoon size limitation. However, individual trucks, often owned by different operators, may prioritize their own gains over the fleet-wide outcome. To maximize fleet-wide benefits while ensuring stable platoon formation and sustained cooperation among heterogeneous trucks, it is essential to address the following key questions:
\vspace{-2pt}
\begin{itemize} 
\item [(i)] What is the optimal coalition structure $\mathcal{P}^*$ of game $G_p$ such that each coalition satisfies the platoon-size constraint $M$ while maximizing the total platooning benefit across the fleet? 
\vspace{2pt}
\item [(ii)] Under the optimal coalition structure $\mathcal{P}^*$, how to allocate the platooning benefits among all participants so that the payoff vector lies in the CS-core? 
\end{itemize}
\vspace{3pt}

\begin{remark}\label{Remark1}
In our previous work~\citep{Bai2025stable}, we showed that when the fleet size satisfies $N\!\leq\!{M}$, the grand coalition constitutes the optimal coalition structure and yields the maximum platooning benefit for all trucks. In contrast, this paper aims to address the payoff allocation problem under platoon-size constraints, i.e., $2\!\leq\!{M}\!<\!N$, to capture a broader and more realistic class of applications. 
\end{remark}

\section{Optimal Platoon Formation}\label{Section 3}
In this section, we provide solutions to problem (i) by developing an efficient method to identify the optimal coalition structure of $G_p$. We first establish conditions that ensure the existence of feasible coalition structures in which every truck can be assigned to a platoon. 
\vspace{3pt}

\begin{lemma}[Feasibility condition] Let $\mathcal{N}$ be a set of $N \ge 2$ players and let $M$ be an integer satisfying $2 \le M < N$. A coalition structure $\mathcal{P}=\{\mathcal{S}_1,\dots,\mathcal{S}_P\}$ that forms a partition of $\mathcal{N}$ and satisfies $2 \le |\mathcal{S}_m| \le M$ for all $m$ exists if and only if there exists an integer $P \ge 1$ such that
\begin{align}
\left\lceil \frac{N}{M} \right\rceil \le P \le \left\lfloor \frac{N}{2} \right\rfloor.
\label{Equ.3}
\end{align}\label{Lemma1}
\end{lemma}
\vspace{-3pt}

\begin{proof}
Recall that $2\!\leq\!{M}\!<\!N$. We prove the necessity and sufficiency of the feasibility condition, respectively.
\vspace{-3pt}

($\Rightarrow$) Necessity.\\
Suppose there exists a feasible coalition structure $\mathcal{P}\!=\!\{\mathcal{S}_1,\dots,\mathcal{S}_P\}$ such that $2\!\le\!|\mathcal{S}_m|\!\le\! M$ for all $m$. Summing the coalition sizes yields
\begin{align}
N=\sum_{m=1}^P\!|\mathcal{S}_m|.\nonumber
\end{align}
Since each coalition satisfies $|\mathcal{S}_m|\!\ge\!2$, we have $N\!\ge\!2P$. In addition, as $|\mathcal{S}_m|\!\le\!M$, we obtain $N\!\le\!MP$. Therefore, $2P\!\le\!N\!\le\!MP$ holds, which is equivalent to
\begin{align}
\left\lceil \frac{N}{M}\right\rceil \le P \le \left\lfloor \frac{N}{2}\right\rfloor.\nonumber
\end{align}
\vspace{-5pt}

($\Leftarrow$) Sufficiency.\\
Conversely, assume that there exists an integer $P\!\ge\!1$ satisfying~\eqref{Equ.3}, i.e., $2P\!\le\!N\!\le\!MP$. Below, we will show that $N$ trucks can be partitioned into $P$ platoons whose sizes lie in $[2,M]$ and sum to $N$. We discuss two cases:
\vspace{-3pt}

\emph{Case (1) $M\!=\!2$.} In this case, $2P\!\le\!N\!\le\!2P$ forces $N\!=\!2P$. Thus, the only feasible coalition structure consists of $P$ platoons, each of size $2$.
\vspace{-3pt}

\emph{Case (2) $M\!\ge\!3$.} Let us start with $P$ platoons of size $2$, which corresponds to $2P$ trucks in total. The remaining number of trucks is denoted as  
\begin{align}
r=N-2P\ge0.\nonumber
\end{align}
From $N\!\le\!MP$, we obtain $r\!\le\!(M\!-\!2)P$. Next, we distribute these $r$ trucks across the $P$ platoons by adding at most $(M\!-\!2)$ trucks to each. Define that
\begin{align}
q:=\left\lfloor \frac{r}{M\!-\!2}\right\rfloor, \quad s:=r-q(M\!-\!2).\nonumber
\end{align}
Thus, $0\!\le\!s\!<\!M\!-\!2$. We increase the sizes of $q$ platoons from $2$ to $M$ by adding $(M\!-\!2)$ trucks to each. If $s\!>\!0$, increase one additional platoon from size $2$ to $2\!+\!s$, and all other platoons remain at size $2$. This construction yields $q$ platoons of size $M$, one platoon of size $2+s\!\in\!\{3,\dots,M\!-\!1\}$ if $s\!>\!0$, and all remaining platoons of size $2$. Consequently, all platoon sizes lie in $[2,M]$, and the total number of trucks is 
\begin{align}
2P+q(M\!-\!2)+s=2P\!+r=N.\nonumber
\end{align}
Thereby, a feasible partition of $N$ trucks into $P$ platoons exists. Since every platoon contains at least two trucks, a leader can be designated for each. This proves the feasibility condition in Lemma~\ref{Lemma1}. 
\end{proof}

Lemma~\ref{Lemma1} establishes conditions under which a feasible coalition structure exists under the platoon size constraint. In particular, the lower bound on $P$ in condition~\eqref{Equ.3} guarantees that no platoon exceeds the size limit $M$, while the upper bound ensures that every platoon contains at least two trucks so that no truck fails to join a platoon. On this basis, we present the optimal platoon formation structure for the mixed-energy truck platooning game $G_p$ through the following results.
\vspace{2pt}

\begin{theorem}[Optimal platoon formation]\label{Theorem1}
Let $N_e\!=\!|\mathcal{N}_e|$ and $N_f\!=\!|\mathcal{N}_f|$ denote the numbers of ETs and FPTs in $\mathcal{N}$, with $N_e\!+\!N_f\!=\!N$. 
Suppose the feasibility condition in Lemma~\ref{Lemma1} holds. Then:\\
(a) The optimal coalition structure of game $G_p\!=\!(\mathcal{N},v)$, i.e., the one that maximizes the total platooning benefit, is obtained by forming the smallest number of feasible platoons, denoted as $P^*$, which equals
\begin{align}\label{Equ.4}
P^*\!=\!\begin{cases}
\left\lceil \dfrac{N}{M} \right\rceil, & \text{if } M\!\ge\!3,\\[2ex]
\dfrac{N}{2}, & \text{if } M\!=\!2 \text{ and } N \text{ is even}.
\end{cases}
\end{align}
(b) Let $L_e^*$ and $L_f^*$ be the numbers of ET and FPT leaders assigned to the $P^*$ platoons, which are determined by
\begin{align}
L_e^*=\min\{N_e,P^*\}, \quad  L_f^*=P^*\!-L_e^*.\label{Equ.5}
\end{align}
\end{theorem}
\vspace{-3pt}

\begin{proof}
Consider any feasible coalition structure $\mathcal{P}\!=\!\{\mathcal{S}_1,\dots,\mathcal{S}_P\}$ with $2\!\le\!|\mathcal{S}_m|\!\le\!M$, and denote as $L_e$, $L_f$ the numbers of ET and FPT leaders, respectively, satisfying
\begin{align}
L_e+L_f=P, \quad 0\leq{L_e}\leq{N_e}, \quad 0\leq{L_f}\leq{N_f}.\nonumber
\end{align}
The numbers of ET and FPT followers are denoted as
\begin{align}
F_e=N_e\!-\!L_e,\quad F_f=N_f\!-\!L_f.\nonumber
\end{align}
In line with the characteristic function given in~\eqref{Equ.2}, the total platooning benefit under coalition structure $\mathcal{P}$ is
\begin{align}
V(\mathcal{P})&=\sum_{m=1}^P\!v(\mathcal{S}_m)\nonumber\\
&=\epsilon_e(N_e\!-\!L_e)+\epsilon_f(N_f\!-\!L_f)\nonumber\\
&=\underbrace{\big(\epsilon_e N_e\!+\!\epsilon_f N_f\big)}_{\text{Constant in }\mathcal{P}}
-\big(\epsilon_e L_e\!+\!\epsilon_fL_f\big).\label{Equ.6}
\end{align}
Thus, maximizing $V(\mathcal{P})$ over all feasible platoon coalition structures is equivalent to minimizing $(\epsilon_e L_e\!+\epsilon_fL_f)$ subject to the feasibility constraints on $P$, $L_e$ and $L_f$. Below, we prove results (a) and (b), respectively.
\vspace{-3pt}

\noindent(a)
For any fixed $P$, we have 
\begin{align}
\epsilon_e L_e+\epsilon_f L_f\geq{P\cdot\min\{\epsilon_e,\epsilon_f\}}.\nonumber
\end{align}
By~\eqref{Equ.6}, the total benefit $V(\mathcal{P})$ is nonincreasing in $P$. Therefore, maximizing $V(\mathcal{P})$ requires selecting the smallest feasible number of platoons. From Lemma~\ref{Lemma1}, when $M\!=\!2$, a feasible partition exists if and only if $N$ is even. In this case, the number of platoons is uniquely determined by $P\!=\!N/2$. When $M\!\geq\!{3}$, as the interval $\big[\lceil N/M\rceil,\lfloor N/2\rfloor\big]$ is nonempty, a feasible partition exists for every $N$, and the minimum feasible number of platoons is
\begin{align}
P_{\min}=\left\lceil \frac{N}{M}\right\rceil.\label{Equ.7}
\end{align}
This proves \eqref{Equ.4} in part (a).
\vspace{-3pt}

\noindent(b) Under the optimal coalition structure $\mathcal{P}^*$, $P\!=\!P^*$ is fixed. By~\eqref{Equ.6}, maximizing $V(\mathcal{P})$ is equivalent to solving
\begin{align}
\min_{L_e, L_f} \ \ &\epsilon_e L_e\!+\!\epsilon_f L_f\nonumber\\
\text{s. t.} \ \ & L_e\!+\!L_f=P^*, \ 0\leq\!{L_e}\!\le{N_e},\ 0\leq\!L_f\!\leq{N_f}.\nonumber
\end{align}
Since $\epsilon_e\!\le\!\epsilon_f$, and 
\begin{align}
\epsilon_eL_e+\epsilon_fL_f=\epsilon_fP^*\!+\!(\epsilon_e\!-\!\epsilon_f)L_e,\nonumber
\end{align}
the objective is minimized by choosing the maximum possible $L_e$. Thus, $L_e^*\!=\!\min\{N_e,P^*\}$. The remaining platoons must then be led by FPTs, so $L_f^*\!=\!P^*\!-\!L_e^{*}$. This establishes part (b) and completes the proof.
\end{proof}

Theorem~\ref{Theorem1} indicates that the total platooning benefit is maximized when the fleet forms the minimum feasible number of platoons. This follows from the fact that each additional platoon introduces an extra leader who contributes no benefit and reduces the number of followers generating positive gains. Notice that any coalition structure satisfying both conditions (a) and (b) of Theorem~\ref{Theorem1} constitutes an optimal platoon formation $\mathcal{P}^*$, which may not be unique. For any feasible problem setting, such an optimal formation can be constructed by Algorithm~\ref{Algorithm1}.

\begin{algorithm}[t]
  \caption{Construct the Optimal Platoon Formation}
  \label{Algorithm1}
  \begin{algorithmic}[1]
  
    \Statex \textbf{Input:} Truck set $\mathcal{N}\!=\!\mathcal{N}_e\!\cup\!\mathcal{N}_f$; platoon size limit $N\!>\! M\! \ge\! 2$ satisfying Lemma~\ref{Lemma1}.
    \Statex \textbf{Output:} Optimal platoon formation structure $\mathcal{P}^*$.

    \State Compute $P^*$, $L_e^*$, and $L_f^*$ according to Theorem~\ref{Theorem1}.
    \State Select $L_e^*$ trucks from $\mathcal{N}_e$ and $L_f^*$ trucks from $\mathcal{N}_f$ to form 
           the leader set $\mathcal{L}^*\!=\!\{L_1,\ldots,L_{P^*}\}$.
    \State Initialize $\mathcal{P}^* \gets \emptyset$ and $\mathcal{F} \gets \mathcal{N}\!\setminus\!\mathcal{L}^*$.

    \For{$m=1$ to $P^*\!-\!1$}
        \State Create platoon $\mathcal{S}_m \gets \{L_m\}$.

        \While{$|\mathcal{S}_m|\!<\!M$}
            \State Select an arbitrary follower $i\!\in\!\mathcal{F}$.
            \State $\mathcal{S}_m \gets \mathcal{S}_m\!\cup\!\{i\}$; \quad
                   $\mathcal{F} \gets \mathcal{F}\!\setminus\!\{i\}$.
        \EndWhile
        \State $\mathcal{P}^* \gets \mathcal{P}^*\!\cup\!\{\mathcal{S}_m\}$.
    \EndFor

    \State Create the last platoon $\mathcal{S}_{P^*} \gets \{L_{P^*}\}$.

    \While{$|\mathcal{S}_{P^*}|\!<\!M$ \textbf{and} $|\mathcal{F}|\!>\!0$}
        \State Select an arbitrary follower $i\!\in\!\mathcal{F}$.
        \State $\mathcal{S}_{P^*} \gets \mathcal{S}_{P^*}\!\cup\!\{i\}$; \quad
               $\mathcal{F} \gets \mathcal{F}\!\setminus\!\{i\}$.
    \EndWhile
    \If{$|\mathcal{S}_{P^*}|\!=\!1$}
        \State Select any index $q\!\in\!\{1,\ldots,P^*-1\}$ with $|\mathcal{S}_q|\!>\!2$.
        \State Select an arbitrary follower $j \in \mathcal{S}_q \!\setminus\!\{L_q\}$.
        \State $\mathcal{S}_{P^*} \gets \mathcal{S}_{P^*}\!\cup\!\{j\}$; \quad
               $\mathcal{S}_q \gets \mathcal{S}_q\!\setminus\!\{j\}$.
    \EndIf

    \State $\mathcal{P}^* \gets \mathcal{P}^*\!\cup\!\{\mathcal{S}_{P^*}\}$.
    \State \Return $\mathcal{P}^*$.

  \end{algorithmic}
\end{algorithm}

\section{Type-based Least-core Payoff Allocation Scheme}\label{Section 4}
In this section, we proceed to address question (ii) raised in Section~\ref{Section 2}. While the optimal coalition structure $\mathcal{P}^*$ maximizes the fleet-wide platooning benefit, trucks owned by different operators may prioritize their individual gains. An unstable allocation could incentivize trucks to leave their assigned platoons or form alternative coalitions, thereby reducing the realized benefit and undermining the sustainability of the optimal formation. To avoid such deviations, we develop a type-based payoff allocation scheme under $\mathcal{P}^*$ that achieves maximum coalition stability. 

Building on the CS-core of a general coalitional game given in Definition~\ref{Def4}, we now specialize this concept to the platoon formation game $G_p$ and introduce the notion of the least-core. 

\begin{definition}[CS-core of $G_p$]\label{Def5}
Consider the game $G_p\!=\!(\mathcal{N},v)$ under the coalition structure $\mathcal{P}^*\!=\!\{\mathcal{S}_1,\dots,\mathcal{S}_{P^*}\}$ determined by Theorem~\ref{Theorem1}. A payoff vector $x\!=\![x_1,\dots,x_N]^{\top}\!\in\!\mathbb{R}^N$ is said to be feasible under $\mathcal{P}^*$ if
\begin{align}
\sum_{i\in\mathcal{S}_m}\!\!x_i=v(\mathcal{S}_m), \quad \forall{\mathcal{S}_m}\!\in\!\mathcal{P}^*.\label{Equ.8}
\end{align}
The CS-core of $G_p$ associated with $\mathcal{P}^*$ is defined as
\begin{align}
&\mathrm{CS\text{-}core}(\mathcal{P}^*)\nonumber\\
&\!=\!\Big\{x\!\in\!\mathbb{R}^N \big| \!\sum_{i\in\mathcal{S}_m}\!x_i\!=\!v(\mathcal{S}_m), \forall{\mathcal{S}_m}\!\in\!\mathcal{P}^*, \text{ and} \nonumber\\
&\quad \quad \quad \quad \quad \sum_{i\in\mathcal{C}}x_i\!\ge\!v(\mathcal{C}), \forall\mathcal{C}\!\subseteq\!\mathcal{N},\text{with } 2\!\le\!|\mathcal{C}|\!\le\!M\Big\}.\label{Equ.9}
\end{align}
\end{definition}

\begin{definition}[Least-core of $G_p$]\label{Def6}
Consider the game $G_p\!=\!(\mathcal{N},v)$ under the coalition structure $\mathcal{P}^*\!=\!\{\mathcal{S}_1,\dots,\mathcal{S}_{P^*}\}$ determined by Theorem~\ref{Theorem1}. A payoff vector $x\!=\![x_1,\dots,x_N]^{\top}\!\in\!\mathbb{R}^N$ is said to be $\varepsilon$--feasible if 
\begin{subequations}\label{Equ.10}
\begin{align}
&\sum_{i\in\mathcal{S}_m}\!x_i= v(\mathcal{S}_m), \quad \forall\mathcal{S}_m\!\in\!\mathcal{P}^*, \label{Equ.10a}\\
&\sum_{i\in\mathcal{C}}x_i\ge v^*(\mathcal{C})-\varepsilon, \quad \forall\mathcal{C}\!\subseteq\!\mathcal{N}, \text{with } 2\!\le\!|\mathcal{C}|\!\le\!{M},\label{Equ.10b}
\end{align}
\end{subequations}
where $v^*(\mathcal{C})$ denotes the maximal value attainable by $\mathcal{C}$, as given in~\eqref{Equ.2}. The least-core radius of $G_p$ under $\mathcal{P}^*$ is then defined as
\begin{equation}\label{Equ.11}
\begin{aligned}
\varepsilon^*&=\min_{x\in\mathbb{R}^N,\ \varepsilon\geq{0}} \ \varepsilon\\
&\mathrm{s.\,t.} \ \ \text{ $x$ is $\varepsilon$--feasible}.
\end{aligned}
\end{equation}
Accordingly, the set 
\begin{align}\label{Equ.12}
\mathrm{LC}(G_p,\mathcal{P}^*)=\big\{x\!\in\!\mathbb{R}^N |\ x \text{ is } \varepsilon^*\text{--feasible}\big\}
\end{align}
is known as the least-core of $G_p$ associated with $\mathcal{P}^*$.
\end{definition}
\vspace{2pt}

\begin{remark}\label{Remark2}
The quantity $\varepsilon^*$ represents the minimal uniform relaxation of the coalition-stability constraints required to sustain $\mathcal{P}^*$ against all admissible deviations. If $\varepsilon^* = 0$, then the CS-core of $G_p$ under $\mathcal{P}^*$ is nonempty. If $\varepsilon^* > 0$, the value of $\varepsilon^*$ quantifies the least amount of relaxation needed to achieve approximate stability, with smaller values indicating a higher degree of stability for the optimal platoon structure.
\end{remark}

In what follows, we present a type-based payoff allocation scheme under the optimal platoon formation $\mathcal{P}^*$, where each ET and FPT receives a uniform payoff $x_e$ and $x_f$, respectively. To handle both CS-core stable and unstable scenarios within a unified framework, the scheme is designed to maximize the achievable CS-core stability. 

Let $\mathcal{P}^*$ be the optimal coalition structure of $G_p$ determined by Theorem~\ref{Theorem1}. Consider the type-based payoff allocations of the following form:
\begin{align}\label{Equ.13}
x_i =
\begin{cases}
x_e, & \text{if $T_i\!=\!e$},\\
x_f, & \text{if $T_i\!=\!f$}.
\end{cases}
\end{align}
\begin{proposition}\label{Proposition1}
Define $(x_e^*,x_f^*,\varepsilon^*)$ as an optimal solution of the following problem:
\begin{align}
\min_{x_e,x_f,\varepsilon \ge 0}  
& \quad \varepsilon \tag{14} \\[4pt]
\mathrm{s.\ t.}\quad
& x_e N_e\!+\!x_fN_f=\epsilon_e(N_e\!-\!L_e^*)\!+\!\epsilon_f(N_f\!-\!L_f^*), 
        \tag{14a} \\
& x_eC_e\!+\!x_f C_f\ge v^*(\mathcal{C})-\varepsilon,\nonumber\\
   & \quad \quad \quad \quad \quad \quad \quad \forall \mathcal{C}\!\subseteq\!\mathcal{N},~
    2\!\le\!|\mathcal{C}|\!\le\!M, \tag{14b} \\
& x_e, x_f\!\in\!\mathbb{R}_+, \tag{14c}
\end{align}
where $C_e$ and $C_f$ denote the numbers of ETs and FPTs in coalition $\mathcal{C}$ with $C_e\!+\!C_f\!=\!|\mathcal{C}|$. Let $x^*\!\in\!\mathbb{R}^N$ be the induced payoff vector, with $x_i^*\!=\!x_e^*$ and $x_i^*\!=\!x_f^*$. Then:\\
(a) If $\varepsilon^*\!=\!0$, then $x^*$ lies in the CS-core of 
      $G_p$ under $\mathcal{P}^*$.\\
(b) If $\varepsilon^*\!>\!0$, $x^*$ minimizes the least-core radius over all allocations of the form \eqref{Equ.13}.
\end{proposition}
\vspace{-3pt}

\begin{proof}
The efficiency constraint (14a) ensures that $x$ is feasible under $\mathcal{P}^*$. 
If $\varepsilon^*\!=\!0$, all stability constraints satisfy
\begin{align}
x_e^*C_e+x_f^*C_f\geq{v^*(\mathcal{C})},\quad \forall\mathcal{C}\!\subseteq\!\mathcal{N},~
    2\!\le\!|\mathcal{C}|\!\le\!M.\nonumber
\end{align}
Thus, $x^*\in\mathrm{CS}$-$\mathrm{core}(\mathcal{P}^*)$ by Definition~\ref{Def4}.  
If $\varepsilon^*\!>\!0$, solving the optimization problem (14) minimizes the maximal stability violation over all type-based allocations. By the definition of the least-core radius, $x^*$ attains the 
minimum least-core radius among all allocations of the form \eqref{Equ.13}.
\end{proof}

Note that, the optimization problem in Proposition~\ref{Proposition1} is a small linear
program (LP) in decision variables $(x_e,x_f,\varepsilon)$ and a set of
linear constraints indexed by all admissible coalitions $\mathcal{C}$, $2 \!\le\!|\mathcal{C}|\!\le\!M$. Since $v^*(\mathcal{C})$ admits the closed-form in \eqref{Equ.2} and every constraint $x_eC_e\!+\!x_fC_f\!\ge\!v^*(\mathcal{C})\!-\!\varepsilon$ is linear in $(x_e,x_f,\varepsilon)$, the entire problem can be solved efficiently using standard LP solvers.

\begin{remark}
Although the number of feasible coalitions $\mathcal{C}$ grows combinatorially
with $N$, the type symmetry of the platoon formation game reduces the effective
constraint set to only $O(M^2)$ candidate coalitions, each characterized by a
pair $(C_e,C_f)$ with $C_e\!+\!C_f\!\le\!M$. As a result, the LP remains small and can be solved efficiently even for large-scale mixed-energy truck fleets.
\end{remark}
\vspace{1pt}

\begin{proposition}
The optimization problem in Proposition~\ref{Proposition1} is always feasible and admits an optimal solution.
\end{proposition}
\vspace{-3pt}
\begin{proof} Since $N_e\!+\!N_f\!=\!N\!\ge\!2$ and all values $v^*(\mathcal{C})$ are finite, the total platooning value under $\mathcal{P}^*$ is denoted as
\begin{align}
V_{\mathrm{tot}}=\epsilon_e (N_e\!-\!L_e^*)\!+\!\epsilon_f(N_f\!-\!L_f^*).\nonumber
\end{align}
Consider the efficiency constraint 
\begin{align}
x_e N_e + x_f N_f = V_{\mathrm{tot}}.\nonumber
\end{align}
Then, a feasible type-based allocation $(x_e,x_f)$ can always be constructed with nonnegative components, and one such construction is 
\begin{align}
(x_e^{(0)},x_f^{(0)}) =
\begin{cases}
\!\big(0, V_{\mathrm{tot}}/N_f\big), & \text{if } N_f\!>\!0,\\[0.5ex]
\!\big(V_{\mathrm{tot}}/N_e, 0\big), & \text{if } N_f\!=\!0.\nonumber
\end{cases}
\end{align}
Given $(x_e^{(0)},x_f^{(0)})$, let us define
\begin{align}
\varepsilon^{(0)} 
\!:= \max\Big\{0,\max_{\mathcal{C}\subseteq\mathcal{N}, 2\le|\mathcal{C}|\le{M}}
\big[v^*(\mathcal{C})\!-\!x_e^{(0)}C_e\!-\!x_f^{(0)}C_f\big]\Big\}.\nonumber
\end{align}
By this construction, $(x_e^{(0)},x_f^{(0)},\varepsilon^{(0)})$ satisfies all the constraints of Proposition~\ref{Proposition1} and thus is a feasible solution of the problem. In addition, since the objective is to minimize $\varepsilon \!\ge\!0$ over a nonempty, closed, and bounded feasible set, the optimal value $\varepsilon^*$ is finite, and an optimal solution $(x_e^*,x_f^*,\varepsilon^*)$ exists for any given $G_p$.
\end{proof}

\section{Numerical Study}\label{Section 5}
This section presents numerical studies to validate the optimality of the proposed platoon formation structure and the stability performance of the type-based payoff allocation scheme. We first verify the optimality of the developed platoon formation with an illustrative example, subject to varying platoon size limitations. Subsequently, we conduct extensive numerical variations and comparisons against multiple baseline methods to further evaluate the stability of the proposed payoff allocation scheme.

\subsection{Optimality of the Platoon Formation}
We consider a mixed-energy truck fleet at a hub with the truck set $\mathcal{N}\!=\!\{1,\dots,9\}$, where the ET set is $\mathcal{N}_e\!=\!\{1,2,3\}$ and the FPT set is $\mathcal{N}_f\!=\!\{4,\dots,9\}$. The platoon size is constrained by $M\!=\!4$. Following the realistic platooning benefit evaluation~\citep{10932686}, the parameters are set as $\epsilon_f\!=\!0.07$ and $\epsilon_e\!=\!0.048$ dollars per kilometer. 

\begin{figure}
\begin{center}
\includegraphics[width=8.8cm]{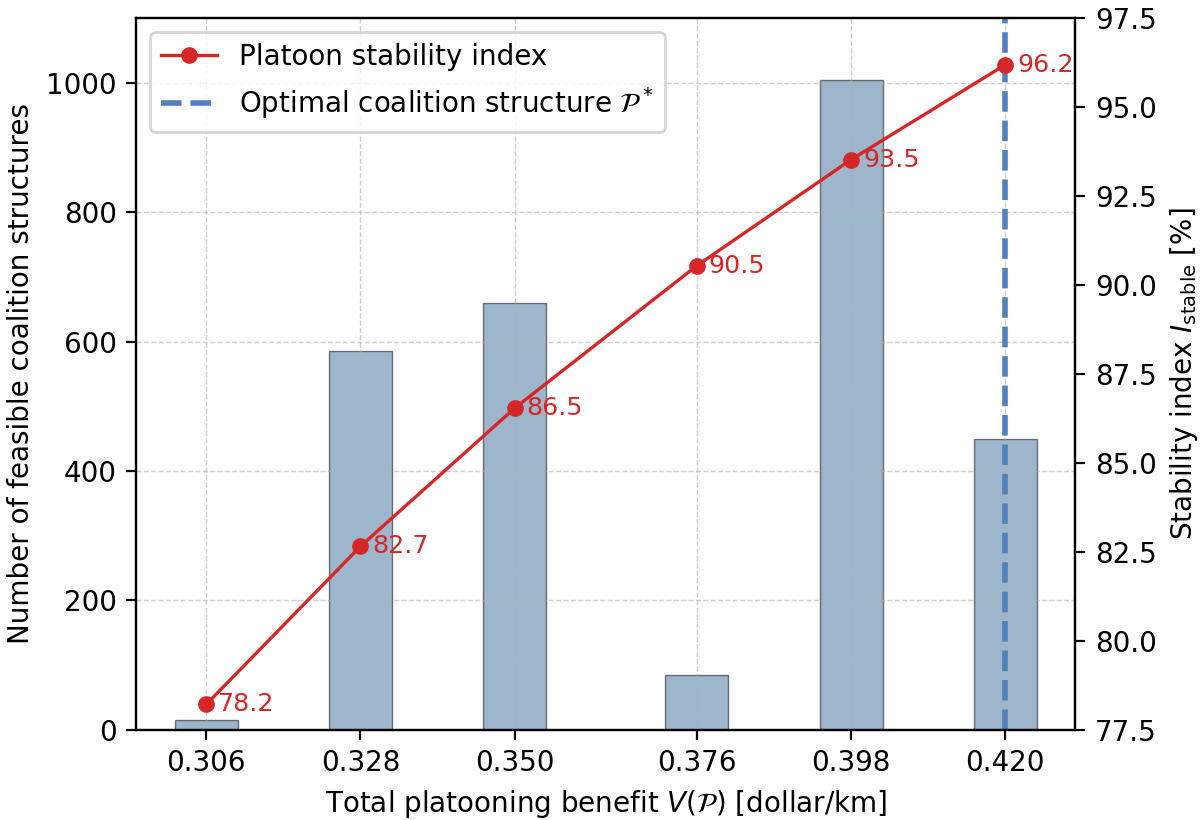}
\caption{Comparison of the total platooning benefit and stability index in different feasible coalition structures.} 
\label{Fig.1}
\end{center}
\end{figure}

As the feasibility condition in Lemma~\ref{Lemma1} is satisfied, we proceed to construct the optimal platoon structure. Applying Theorem~\ref{Theorem1} and Algorithm~\ref{Algorithm1}, we have
\begin{align}
P^*\!=\!\left\lceil \dfrac{N}{M} \right\rceil\!=\!\left\lceil \dfrac{9}{4} \right\rceil\!=\!3, \quad L_e^*\!=\!3, \quad  L_f^*\!=\!0.\nonumber
\end{align}
Thus, an optimal platoon coalition that maximizes the total platooning benefit of the fleet is obtained as 
\begin{align}
\mathcal{P}^*=\big\{\{1,4,5,6\}, \{2,7,8\}, \{3,9\}\big\}.\nonumber
\end{align}
The total benefit achieved by the fleet under $\mathcal{P}^*$ is 
\begin{align}
V(\mathcal{P}^*)=6\epsilon_f=0.42~\mathrm{dollar/km}.\nonumber
\end{align}
To demonstrate the optimality of $\mathcal{P}^*$, we show in Fig.~\ref{Fig.1} the total platooning benefit attained by all feasible coalition structures. The x-axis represents the total benefit of each feasible coalition, while the y-axis shows the number of coalition structures achieving the corresponding benefit, since different truck groupings may yield the same total benefits. In this example, $6$ feasible coalition structures exist, including $\mathcal{P}^*$. The other coalition structures are
\begin{align}
\mathcal{P}_1&=\big\{\{1,2,3\}, \{4,5\}, \{6,7\}, \{8,9\}\big\},\nonumber\\[0.5ex]
\mathcal{P}_2&=\big\{\{1,2\}, \{3,4\}, \{5,6\}, \{7,8,9\}\big\},\nonumber\\[0.5ex]
\mathcal{P}_3&=\big\{\{1,4\}, \{2,5\}, \{3,6\}, \{7,8,9\}\big\},\nonumber\\[0.5ex]
\mathcal{P}_4&=\big\{\{1,2,3\}, \{4,5\}, \{6,7,8,9\}\big\},\nonumber\\[0.5ex]
\mathcal{P}_5&=\big\{\{1,2\}, \{3,4,5\}, \{6,7,8,9\}\big\}.\nonumber
\end{align}
The corresponding total platooning benefits are given by
\begin{align}
V(\mathcal{P}_1)&=0.306, \quad  V(\mathcal{P}_2)=0.328, \quad V(\mathcal{P}_3)=0.350,\nonumber\\[0.5ex]
V(\mathcal{P}_4)&=0.376, \quad V(\mathcal{P}_5)=0.398.\nonumber
\end{align}
The results show that the proposed platoon formation structure $\mathcal{P}^*$ achieves the largest total platooning benefit across all feasible configurations, thereby validating the optimality of the proposed formation approach.

To further evaluate the stability of each feasible coalition structure $\mathcal{P}$ under the type-based least-core allocation, we introduce the following \textit{platoon stability index}:
\begin{align}\tag{15}
I_{\text{stable}}(\mathcal{P},x):=\left(1-\frac{\varepsilon^*(x)}{V(\mathcal{P})}\right)\!\times\!{100\%},
\end{align}
where $V(\mathcal{P})\!=\!\sum_{\mathcal{S}_m\in\mathcal{P}}\!v(\mathcal{S}_m)$ denotes the total platooning benefit of $\mathcal{P}$, and $\varepsilon^*(x)$ is the least-core radius corresponding to the optimal allocation, representing the minimum relaxation required for the CS-core of $G_p$ under $\mathcal{P}$ to be nonempty. Under this metric, $I_{\text{stable}}\!=\!100\%$ indicates perfect CS-core stability, whereas smaller values imply increasing incentives for deviation. 

By solving the problem in Proposition~\ref{Proposition1}, the optimal type-based least-core payoff allocation under $\mathcal{P}^*$ is 
\begin{align}
x_e^*\approx0.032, \quad x_f^*\approx0.054,\nonumber
\end{align}
with the least-core radius $\varepsilon^*(x^*)\!\approx\!{0.016}$. The resulting stability indices are shown on the right y-axis of Fig.~\ref{Fig.1}. As shown, the proposed allocation scheme under the optimal coalition structure $\mathcal{P}^*$ attains the highest stability index among all other feasible coalition structures, demonstrating its superior stability performance. 

\begin{figure}
\begin{center}
\includegraphics[width=8.5cm]{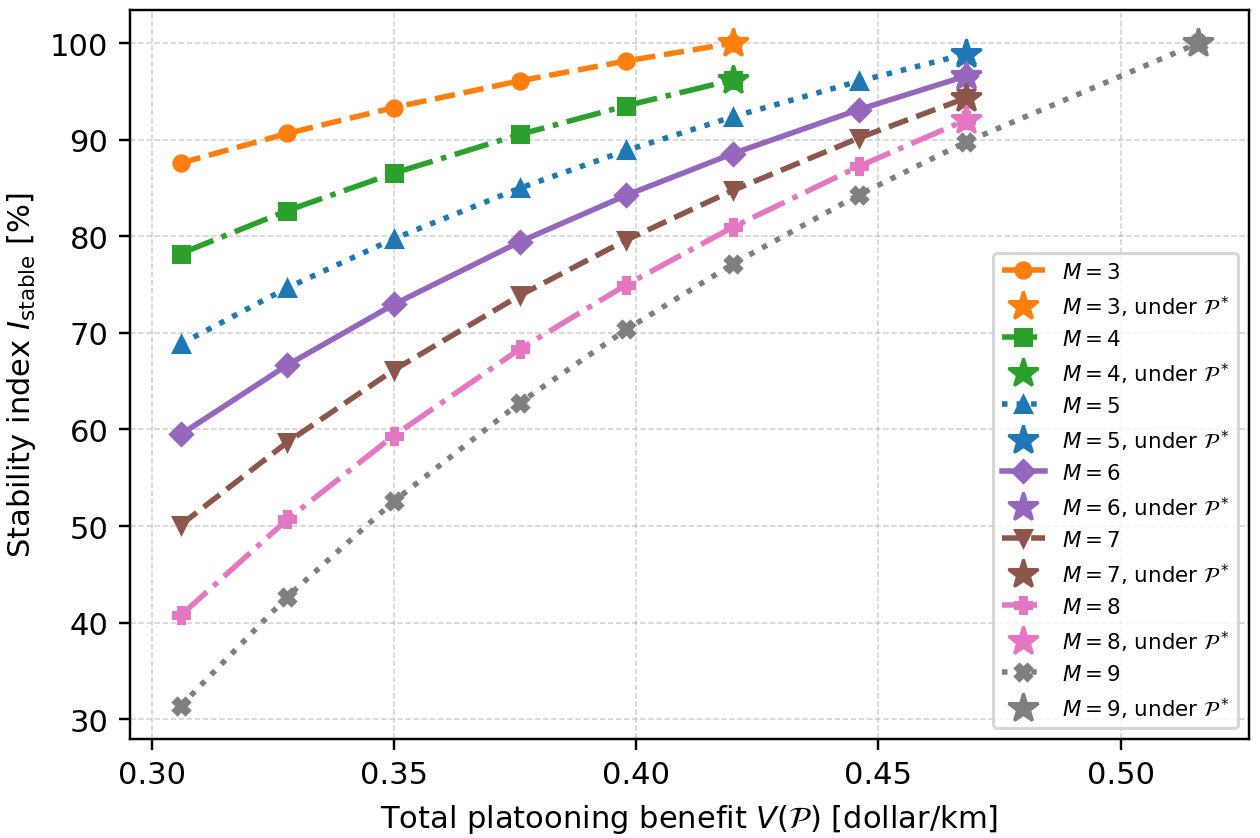}  
\caption{Comparison of the total platooning benefit and stability index under different platoon size limits.} 
\label{Fig.2}
\end{center}
\end{figure}

Fig.~\ref{Fig.2} shows the stability indices of all feasible coalition structures under different platoon size limits $M\!=\!3,\dots,9$, where the stability index of $\mathcal{P}^*$ in each scenario is highlighted with a star. The results show that as $M$ increases, the number of feasible coalition structures grows, while the stability index trends decrease for coalition structures achieving the same total benefit. This is because a larger feasible set introduces more opportunities to form alternative coalitions that might incur higher payoffs, thus increasing trucks' incentives to deviate. Nevertheless, the derived coalition structure $\mathcal{P}^*$ consistently outperforms other coalition structures across all scenarios by achieving the highest total benefit and the highest stability index, demonstrating its optimality and strong stability.

\subsection{Comparison with Baseline Allocation Schemes}
\begin{figure}
\begin{center}
\includegraphics[width=8.85cm]{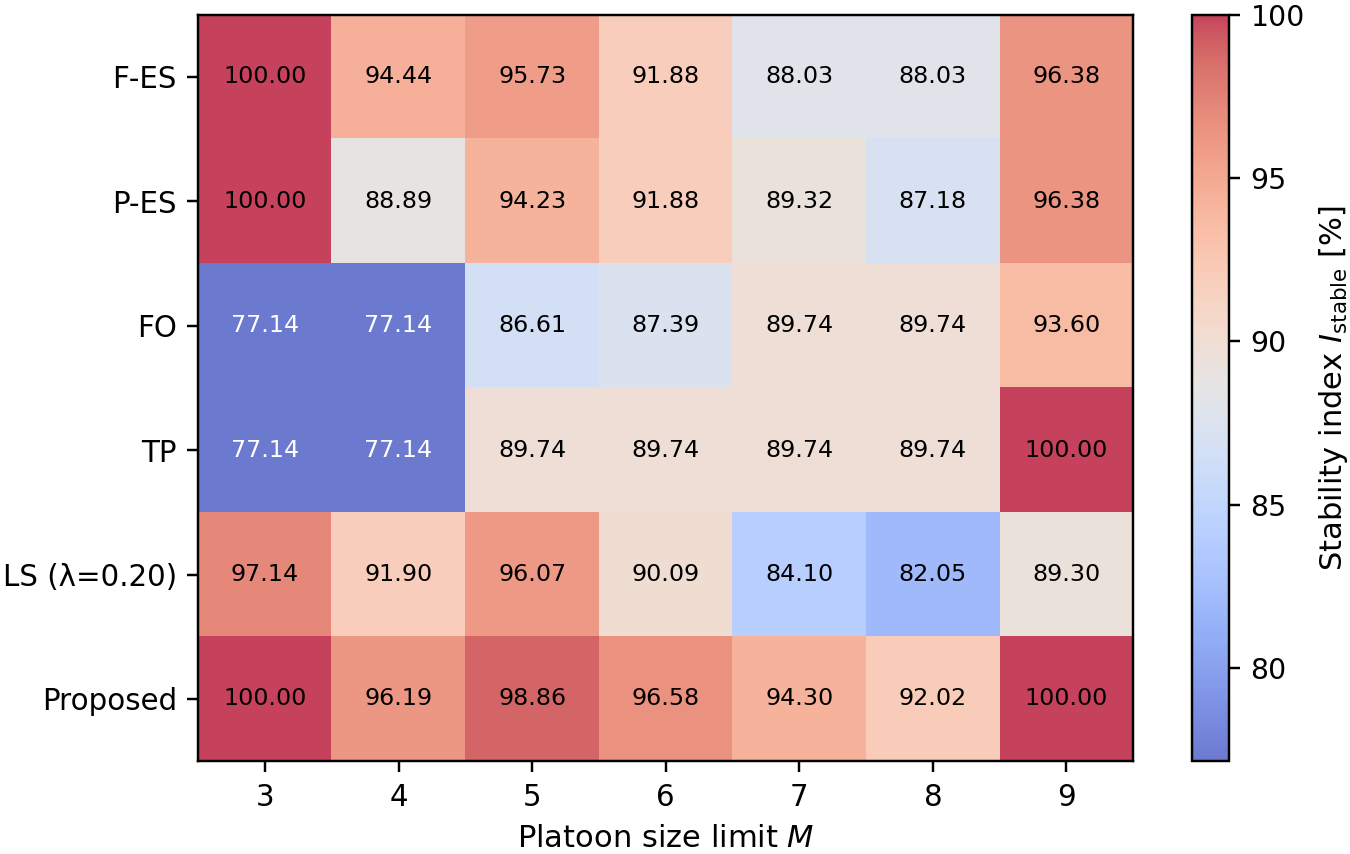}    
\caption{Comparison of the stability index in different payoff allocation schemes under $\mathcal{P}^*$ with different size limits.} 
\label{Fig.3}
\end{center}
\end{figure}

\begin{figure*}[t]
    \centering
    \includegraphics[width=0.99\textwidth]{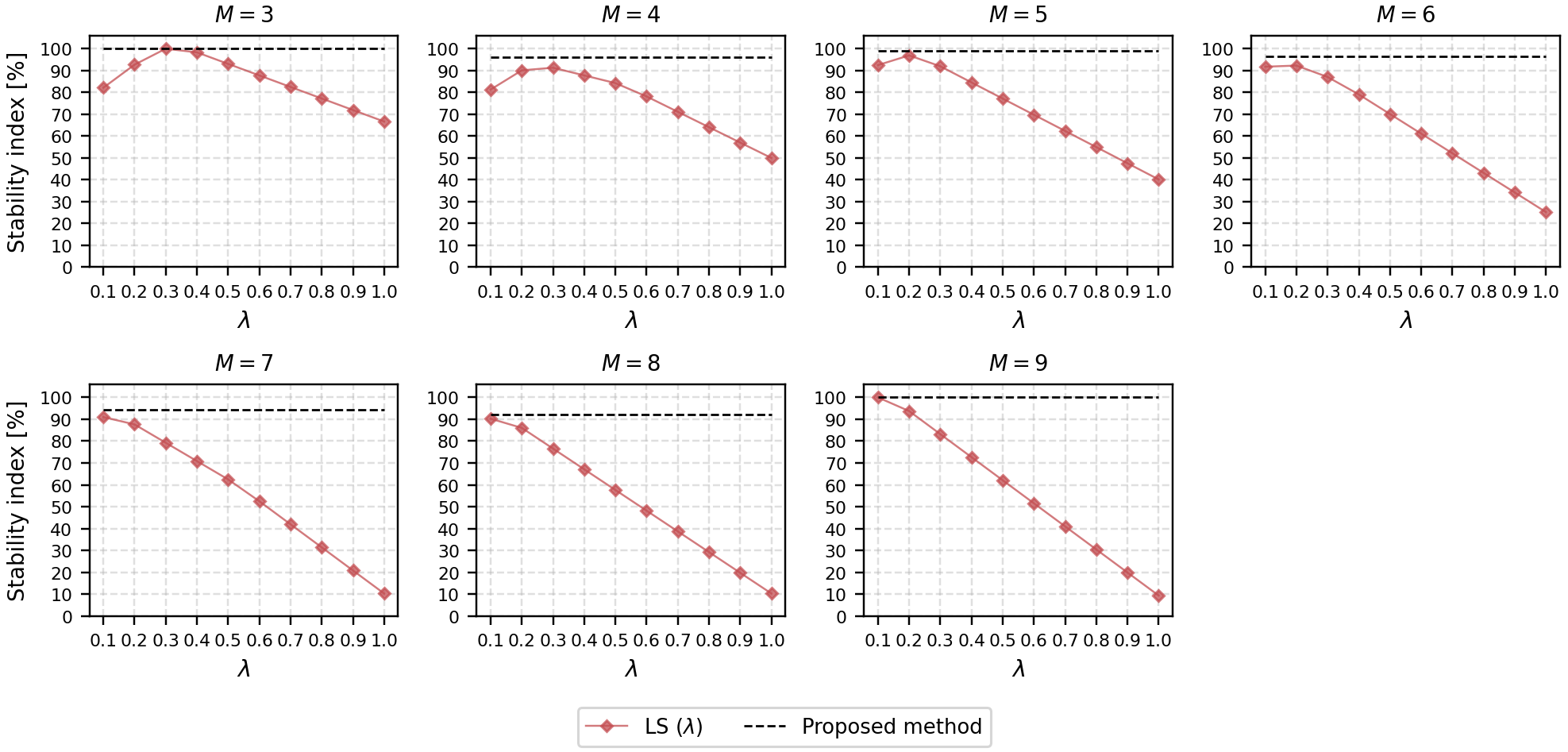}
    \caption{Stability index comparison between the leader–subsidy method (for different values of $\lambda$) and the proposed type-based least-core allocation, under the optimal coalition structure $\mathcal{P}^*$ with $M\!=\!3,\dots,9$.}
    \label{Fig.4}
\end{figure*}

Below, we evaluate the stability of the proposed payoff allocation scheme by comparing it against several baseline allocation methods under the optimal coalition structure $\mathcal{P}^*$. Using the stability index defined in (15), we consider the following baseline approaches for comparison:
\begin{itemize}
\item [\small{$\bullet$}]\textbf{Fleet-based Equal Split (F-ES):}
Each truck in the fleet equally shares the total platooning benefit $V(\mathcal{P}^*)$, regardless of its type or role. 
\vspace{2.5pt}

\item [\small{$\bullet$}]\textbf{Platoon-based Equal Split (P-ES):}
All trucks in each platoon $\mathcal{S}_m\!\in\!\mathcal{P}^*$ receive an equal share of the benefit $v(\mathcal{S}_m)$, independent of their type or role.  
\vspace{2.5pt}

\item [\small{$\bullet$}] \textbf{Follower-only (FO):}
Only followers in each platoon $\mathcal{S}_m\!\in\!\mathcal{P}^*$ share the total benefit $v(\mathcal{S}_m)$ equally, while the platoon leader receives no benefit. 
\vspace{2.5pt}

\item [\small{$\bullet$}] \textbf{Type-proportional (TP):}
Followers share the benefit of each platoon proportionally to their type coefficients $\epsilon_i$, and the leader receives zero. For any truck $i\!\in\!\mathcal{S}_m\!\in\!\mathcal{P}^*$, the allocation is
\begin{align}
x_i=\begin{cases}
0, & \text{if } i\!=\!L_m,\\[4pt]
v(\mathcal{S}_m)\dfrac{\epsilon_{T_i}}{\sum_{j\in\mathcal{S}_m\setminus\{L_m\}}\!\epsilon_{T_j}}, & \text{otherwise.}\nonumber
\end{cases}
\end{align}
\vspace{2.5pt}

\item [\small{$\bullet$}] \textbf{Leader-Subsidy (LS):} The leader of each platoon receives a subsidy $\lambda{v(\mathcal{S}_m)}$ with $\lambda\!\in\!(0,1]$ while platoon followers share the remaining benefit proportionally to their type coefficients. For any truck $i\!\in\!\mathcal{S}_m\!\in\!\mathcal{P}^*$, the allocation is
\begin{align}
x_i=\begin{cases}
\lambda\,v(\mathcal{S}_m), &\text{if } i\!=\!L_m,\\[4pt]
(1\!-\!\lambda)\dfrac{v(\mathcal{S}_m)\epsilon_{T_i}}{\sum_{j\in\mathcal{S}_m\setminus\{L_m\}}\!\epsilon_{T_j}}, & \text{otherwise}.
\end{cases}\nonumber
\end{align}
\end{itemize}
\vspace{2pt}

Fig.~\ref{Fig.3} shows the stability indices of the proposed method and the baseline allocation methods under the optimal coalition structure $\mathcal{P}^*$, with the leader incentive coefficient selected as $\lambda\!=\!0.2$ and the platoon size limit varying as $M\!=\!3,\dots,9$. As shown in the figure, the FO and TP schemes yield lower stability indices across most settings, indicating that assigning zero benefit to leaders undermines coalition stability. The F-ES and P-ES schemes show comparable stability performance, offering moderate stability. While the least-core radius $\varepsilon^*(x)$ of our method is nonzero for some cases (e.g., $M\!=\!4,\dots,8$), i.e., the CS-core stability is not perfect, the proposed type-based least-core allocation scheme consistently achieves the highest stability index in all scenarios, showing good stability and robustness against deviations.

To further evaluate how the subsidy coefficient affects the stability of the LS scheme, Fig.~\ref{Fig.4} shows the stability indices of LS and the proposed method for $\lambda\!=\!0.1,0.2,\dots,1.0$. As we can see, the stability index of the LS scheme is highly sensitive to the choice of $\lambda$. Within a proper range, increasing $\lambda$ contributes to the increase of the coalition stability as leaders receive sufficient incentive to remain in the coalition. However, when $\lambda$ exceeds an optimal level, the stability index declines, since the leaders get a larger portion of the total benefit, causing followers to have a stronger incentive to deviate. In contrast, our proposed method (shown by the dashed line) depends only on the truck type and consistently achieves a higher stability index than the best-performing LS scheme. This demonstrates that the proposed payoff allocation scheme provides a more balanced and robust incentive structure across all trucks. 

Finally, we present in Table~\ref{Table 1} the payoffs received by each ET and FPT by applying the proposed payoff allocation scheme, along with the stability indices. The results show that the stability index remains above $92\%$ across all platoon size limits, indicating a high level of stability. Moreover, both the total platooning benefit and the payoffs allocated to ETs and FPTs increase as $M$ grows. This aligns with the results in our previous work~\citep{Bai2025stable}, i.e., when no platoon size constraint is imposed, forming a grand coalition maximizes the overall platooning benefit and the benefits of individual platooning participants. The above validation and results confirm that $\mathcal{P}^*$ maximizes the fleet’s total platooning benefit, and the proposed type-based least-core allocation achieves superior CS-core stability compared with all baselines.

\begin{table}[t!]\label{Table1}
\centering
\caption{Optimal payoff allocations and stability indices of the proposed approach.}
\label{Table 1}
\renewcommand{\arraystretch}{1.1}
\setlength{\tabcolsep}{4pt}
\begin{tabular}{cccccc}
\hline
~\textbf{$M$} & ~~$\boldsymbol{I_{\mathrm{stable}}}~[$\%$]$ & ~$\boldsymbol{\varepsilon^*(x)}$~~ & ~~$V(\mathcal{P}^*)$~~ & ~~~$\boldsymbol{x_e^*}$~~~~ & ~~~~~$\boldsymbol{x_f^*}$~~ \\ \hline
~3 & 100.00 & 0.0000 & 0.4200 & 0.0393 & 0.0503 \\
~4 & 96.19 & 0.0160 & 0.4200 & 0.0320 & 0.0540 \\
~5 & 98.86 & 0.0053 & 0.4680 & 0.0373 & 0.0593 \\
~6 & 96.58 & 0.0160 & 0.4680 & 0.0373 & 0.0593 \\
~7 & 94.30 & 0.0267 & 0.4680 & 0.0373 & 0.0593 \\
~8 & 92.02 & 0.0373 & 0.4680 & 0.0373 & 0.0593 \\ 
~9 & 100.00 & 0.0000 & 0.5160 & 0.0427 & 0.0647 \\ \hline
\end{tabular}
\end{table}

\section{Conclusion}\label{Section 6}
In this paper, we addressed the platoon formation problem for a mixed-energy truck fleet comprised of both FPTs and ETs while taking into account platoon size constraints. We modeled the strategic interactions among trucks in a coalitional game framework and established conditions to ensure the feasibility and optimality of the coalition structure. On this basis, we developed an efficient algorithm to construct the optimal platoon formation, achieving the maximal fleet-wide platooning benefit. Moreover, a type-based least-core payoff allocation scheme was developed under the optimal platoon formation that maximizes the coalition stability. Extensive numerical studies demonstrated that the proposed platoon formation structure achieves the highest total benefit among all feasible coalition structures, and the developed allocation scheme outperforms existing baseline methods. As for future work, we would extend this framework to multi-hub platoon formation and coordination, enabling applications to mixed-energy truck platooning in large-scale road transportation. 

\bibliography{Ting,IDS_Publications_2025}             

@article{Bai2025stable,
	author = {Bai, Ting and Johansson, Karl Henrik and M{\aa}rtensson, Jonas and Malikopoulos, Andreas A},
	journal = {arXiv preprint arXiv:2507.16923},
	title = {Stable and Fair Benefit Allocation in Mixed-Energy Truck Platooning: {A} Coalitional Game Approach},
	year = {2025}}

@inproceedings{Wang2025programming,
	arxivid = {2505.00847},
	author = {Wang, Ying and Bai, Ting and Malikopoulos, Andreas A.},
	booktitle = {28th IEEE International Conference on Intelligent Transportation Systems (ITSC)},
	title = {Platoon Coordination and Leader Selection in Mixed Transportation Systems via Dynamic Programming},
	year = {2025}}

@article{10932686,
	author = {Bai, Ting and Li, Yuchao and Malikopoulos, Andreas A. and Johansson, Karl H. and M{\aa}rtensson Jonas},
	doi = {10.1109/TITS.2025.3550035},
	journal = {IEEE Transactions on Intelligent Transportation Systems},
	number = {7},
	pages = {10278-10294},
	title = {Distributed Charging Coordination for Electric Trucks under Limited Facilities and Travel Uncertainties},
	volume = {26},
	year = {2025}}

@ARTICLE{7904618,
  author={Nicolaides, Doros and Cebon, David and Miles, John},
  journal={IEEE Systems Journal}, 
  title={Prospects for Electrification of Road Freight}, 
  year={2018},
  volume={12},
  number={2},
  pages={1838-1849}}

@article{link2024rapidly,
  title={Rapidly declining costs of truck batteries and fuel cells enable large-scale road freight electrification},
  author={Link, Steffen and Stephan, Annegret and Speth, Daniel and Pl{\"o}tz, Patrick},
  journal={Nature Energy},
  volume={9},
  number={8},
  pages={1032--1039},
  year={2024},
  publisher={Nature Publishing Group UK London}
}

@article{tsugawa2016review,
  title={A review of truck platooning projects for energy savings},
  author={Tsugawa, Sadayuki and Jeschke, Sabina and Shladover, Steven E},
  journal={IEEE Transactions on Intelligent Vehicles},
  volume={1},
  number={1},
  pages={68--77},
  year={2016},
  publisher={IEEE}
}

@article{axelsson2016safety,
  title={Safety in vehicle platooning: A systematic literature review},
  author={Axelsson, Jakob},
  journal={IEEE Transactions on Intelligent Transportation Systems},
  volume={18},
  number={5},
  pages={1033--1045},
  year={2016},
  publisher={IEEE}
}

@ARTICLE{10147895,
  author={Bai, Ting and Li, Yuchao and Johansson, Karl H. and Mårtensson, Jonas},
  journal={IEEE Control Systems Letters}, 
  title={Rollout-Based Charging Strategy for Electric Trucks With Hours-of-Service Regulations}, 
  year={2023},
  volume={7},
  number={},
  pages={2167-2172},
  doi={10.1109/LCSYS.2023.3285137}}

@inproceedings{meisen2008data,
  title={A data-mining technique for the planning and organization of truck platoons},
  author={Meisen, Philipp and Seidl, Thomas and Henning, Klaus},
  booktitle={Proceedings of the International Conference on Heavy Vehicles},
  pages={19--22},
  year={2008}
}

@article{chen2023cost,
  title={Cost allocation of cooperative autonomous truck platooning: Efficiency and stability analysis},
  author={Chen, Shukai and Wang, Hua and Meng, Qiang},
  journal={Transportation Research Part B: Methodological},
  volume={173},
  pages={119--141},
  year={2023},
  publisher={Elsevier}
}

@article{van2017efficient,
  title={Efficient dynamic programming solution to a platoon coordination merge problem with stochastic travel times},
  author={Van de Hoef, Sebastian and Johansson, Karl H and Dimarogonas, Dimos V},
  journal={IFAC-PapersOnLine},
  volume={50},
  number={1},
  pages={4228--4233},
  year={2017},
  publisher={Elsevier}
}

@inproceedings{farokhi2013game,
  title={A game-theoretic framework for studying truck platooning incentives},
  author={Farokhi, Farhad and Johansson, Karl H},
  booktitle={16th IEEE International Conference on Intelligent Transportation Systems (ITSC)},
  pages={1253--1260},
  year={2013}
}

@inproceedings{johansson2019game,
  title={Game theoretic models for profit-sharing in multi-fleet platoons},
  author={Johansson, Alexander and M{\aa}rtensson, Jonas},
  booktitle={IEEE Intelligent Transportation Systems Conference (ITSC)},
  pages={3019--3024},
  year={2019}
}

@article{sun2019behaviorally,
  title={Behaviorally stable vehicle platooning for energy savings},
  author={Sun, Xiaotong and Yin, Yafeng},
  journal={Transportation Research Part C: Emerging Technologies},
  volume={99},
  pages={37--52},
  year={2019},
  publisher={Elsevier}
}

@ARTICLE{9976309,
  author={Johansson, Alexander and Bai, Ting and Johansson, Karl Henrik and Mårtensson, Jonas},
  journal={IEEE Intelligent Transportation Systems Magazine}, 
  title={Platoon Cooperation Across Carriers: From System Architecture to Coordination}, 
  year={2023},
  volume={15},
  number={3},
  pages={132-144}}

@ARTICLE{10209062,
  author={Bai, Ting and Johansson, Alexander and Johansson, Karl Henrik and M{\aa}rtensson, Jonas},
  journal={IEEE Transactions on Intelligent Transportation Systems}, 
  title={Large-Scale Multi-Fleet Platoon Coordination: A Dynamic Programming Approach}, 
  year={2023},
  volume={24},
  number={12},
  pages={14427-14442}}

@article{van2017fuel,
  title={Fuel-efficient en route formation of truck platoons},
  author={Van De Hoef, Sebastian and Johansson, Karl Henrik and Dimarogonas, Dimos V},
  journal={IEEE Transactions on Intelligent Transportation Systems},
  volume={19},
  number={1},
  pages={102--112},
  year={2017},
  publisher={IEEE}
}
\end{document}